\documentclass[twocolumn]{aastex631}

\newcommand{\target}{{\it J064722.95+031644.6~}}
\newcommand{\starget}{{\it J0647~}}
\newcommand{\rev}[1]{{\bf #1}}
\received{}
\revised{}
\accepted{}

\begin{document}

\title{A Gigantic Mid-Infrared Outburst in an Embedded Class-I Young Stellar Object \target}


\correspondingauthor{Tinggui Wang}
\email{twang@ustc.edu.cn}

\author[0000-0002-1517-6792]{Tinggui Wang}
\affiliation{Deep Space Exploration Laboratory /Department of Astronomy, University of Science and Technology of China, Hefei 230026, People’s Republic of China}
\author{Jiaxun Li}
\affiliation{Department of Astronomy, University of Science and Technology of China, Hefei, 230026, China, Hefei, 230026, People’s Republic of China}
\author{Gregory M.Mace}
\affiliation{Department of Astronomy, The University of Texas, Austin, TX 78712, US}

\author{Tuo Ji}
\affiliation{Key Laboratory for Polar Science, MNR, Polar Research Institute of China, 451 Jinqiao Road, Shanghai, 200136, People’s Republic of China}
\author{Ning Jiang}
\affiliation{Department of Astronomy, University of Science and Technology of China, Hefei, 230026, China, Hefei, 230026, People’s Republic of China}
\author{Qingfeng Zhu}
\affiliation{Deep Space Exploration Laboratory /Department of Astronomy, University of Science and Technology of China, Hefei 230026, People’s Republic of China}
\author{Min Fang}
\affiliation{Purple Mountain Observatory, Chinese Academy of Sciences, 10 Yuanhua Road, Nanjing 210023, People's Republic of China}

\begin{abstract}
We report the serendipitous discovery of a giant mid-infrared (MIR) outburst from a previously unknown source near a star-forming region in the constellation Monoceros. The source gradually brightened by a factor of 5 from 2014 to 2016 before an abrupt rise by a factor of more than 100 in 2017. A total amplitude increase of $>$500 at 4.5~\micron\ has since faded by a factor of about 10. Prior to the outburst it was only detected at wavelengths longer than 1.8~\micron\ in UKIDSS, Spitzer, and Herschel with a spectral energy distribution of a Class I Young Stellar Object (YSO). It has not been detected in recent optical surveys, suggesting that it is deeply embedded. With a minimum distance of 3.5 kpc, the source has a bolometric luminosity of at least 9 $L_\sun$ in the quiescent state and 400 $L_\sun$ at the peak of the eruption. The maximum accretion rate is estimated to be at least a few $10^{-5}$ $M_\sun$ year$^{-1}$. It shares several common properties with another eruptive event, WISE~J142238.82-611553.7: exceptionally large amplitude, featureless near-infrared spectrum with the exception of H2 lines, intermediate eruption duration, an embedded Class I YSO, and a low radiative temperature ($<$600-700 K) in outburst. We interpret that the radiation from the inner accretion disk and young star is obscured and reprocessed by either an inflated outer disk or thick dusty outflow on scales $>$6.5 AU during the outburst.
\end{abstract}



\section{Introduction} \label{sec:intro}

Star formation takes place in dense cores of molecular clouds. The initial collapse of the cloud core leads to a small protostar, which grows into a typical main-sequence star by accreting material from the surrounding envelope via a protoplanetary disk. On the basis of the envelope, disk, and stellar core status, YSOs can be divided into Class 0, I, II, and III. Class 0 represents the earliest stage of their evolution, where a protostar is embedded in a quasi-spherical thick envelope; followed by Class I objects, which possess a disk and envelope; and the gas envelope is largely dispersed in Class II objects, leaving a planetary-forming disk surrounding the central star; Class III represents the last evolution stage of YSOs before it becomes a main sequence star with only a debris dust disk \citep{1987ApJ...312..788A}.
Since the typical accretion rate for T-Tauri stars (TTS) is on the order of $10^{-8 - -9}$ M$_\sun$~yr$^{-1}$, far smaller than the requirement for
the growth of a solar mass star, it was proposed that most of the mass is acquired in episodic accretions \citep{2013AJ....145...94D}. Numerical simulations show that accretion rates may increase by up to 3 orders of magnitude during short episodic accretion phases
\citep{2006ApJ...650..956V, 2010ApJ...719.1896V, 2013MNRAS.433.3256V,2021ApJ...909...31K}. 

Episodic accretion manifests itself as optical and infrared eruptions, which are distinguished from more frequent TTS variations by their much larger amplitude and long duration \citep[see review]{2022arXiv220311257F}. The most extreme eruption types are FU Orionis objects (FUors), with amplitudes up to 8 magnitudes and a duration of one hundred years \citep{1977ApJ...217..693H,1989ESOC...33..233H,1996ARA&A..34..207H}.  
The amplitudes of the outbursts in EXors are less than in FUors on average with a duration on a time scale of several months to a year (\citealp{1996ARA&A..34..207H}). But some EXors brighten up to 6 mags \citep{2007A&A...472..207F} in a few months. These two classes of objects exhibit distinct spectroscopic properties. FUors in outbursts are characterized by an F/G supergiant spectrum, but with broad double peak absorption lines in the optical spectrum and K/M supergiants with CO overtones in the near-infrared spectrum. In contrast, classical EXors exhibit K-or-M dwarf absorption spectra and TTS emission spectra in quiescence and an additional thermal component with a temperature of 1000-4500 K during outburst. Hydrogen, metallic or CO lines are seen in both emission and absorption in the infrared during the outburst \citep{2022arXiv220311257F}. Although EXors are Class II YSOs, it is not yet clear whether FUors resemble Class I or Class II YSOs \citep{2018MNRAS.474.4347C,2011ApJ...730...80M}. The different spectroscopic properties of FUors and EXors can be understood by their different accretion modes. Magnetospheric column accretion works in EXors, while the accretion disk extends to the stellar surface in FUors \citep{2022ApJ...927..144R,2022ApJ...936..152L}. With the advance of large-sky surveys, some eruptive YSOs have been found to fill the gap between FUors and EXors in the amplitude and duration of the outburst \citep{2022arXiv220311257F}. 

However, much of the current knowledge on these eruptions is largely biased against Class 1/0 embedded objects. To date, approximately 30 FUors or FUor-like objects are known, most of which have been discovered at optical wavelengths. Systematic surveys at longer wavelengths provided a different view of eruptive YSOs. \citet{2017MNRAS.465.3039C} found 106 eruptive YSOs in the VISTA Variables in the Via Lactea (VVV) survey of the Galactic plane covering 119 square degrees. The sample showed that eruptions were 10 times more common in Class I objects than in Class II objects within the sample, and the typical duration was between those of FUors and EXors. A JCMT transient survey of eight nearby star-forming regions (SFR) in four years resulted in six sources with a year-long eruption with 40\% higher accretion rates than in the quiescent state, including four Class I and two Class 0 sources. \citep{2021ApJ...920..119L}. The latter authors concluded that episodic accretion plays a minimal role in the mass accumulation of stars.  

Although some optically discovered FUors and EXors are less variable in the mid-infrared than in optical \citep{2007A&A...472..207F,2020ApJ...899..130S,2018ApJ...869..146H}, \cite{2020MNRAS.499.1805L} found an 8 magnitude outburst with intermediate duration in the mid-infrared of a previously unknown YSO. PTF 14ig displayed an outburst more than 6 mags in the mid-infrared accompanying with an optical outburst with a similar amplitude\citep{2019ApJ...874...82H}. A similar amplitude eruption was also detected in WISEA J075915.26-310844.6, but its nature is less clear (\citealp{2020RNAAS...4..242T}; see also remarks by Hillenbrand\footnote{https://sites.astro.caltech.edu/$\sim$lah/puppisneb/}).

In this paper, we report the serendipitous discovery of a gigantic eruption in a previously unknown YSO (RA = 06:47:22.95, DEC = +03:16:44.56) in the mid-infrared. The source was identified in the blind search for large amplitude mid-infrared variables in Wide Infrared Sky Explorer (WISE) archive. The paper is arranged as follows. We describe the observations and archival data, as well as the data analysis methods, in \S 2. We present the main results in \S 3. A discussion of the nature of the source and the outburst is given in \S 4. In \S 5 we summarize our conclusions.

\section{Archival Data, New Observations, and Analysis} \label{sec:observations}

Initially, we examined mid-infrared light curves from the W1 and W2 bands of the AllWISE and NEOWISE single-exposure photometric database in the IRSA infrared archive within 6\arcsec~ of the ALLWISE position. Light curves covered the period from early 2010 to the end of 2022 \citep{https://doi.org/10.26131/irsa1,https://doi.org/10.26131/irsa144,2010AJ....140.1868W,2011ApJ...731...53M,2014ApJ...792...30M}. We examined the single-exposure photometric data in each epoch and found that they were consistent, with the exception of the highest flux state in the W2 band when the source was brighter than 7.0 mag in W2, leading to saturation. Therefore, we averaged the single-exposure measurements at each epoch. We noticed that the location of the erupting source was close to an optically bright star (hereafter referred to as S1, 06:47:22.98+03:16:39.9) in Pan-STARRS before mid-2014, but shifted northward from S1 by 4$\farcs$5 since the giant eruption. This indicates that the outburst comes from another source (06:47:22.95+03:16:44.56) that was not detected by Pan-STARRS, rather than due to a large proper motion.  

\begin{figure*}
\figurenum{1}
\centering
\begin{minipage}{0.95\textwidth}
	\centerline{\includegraphics[width=1.0\textwidth]{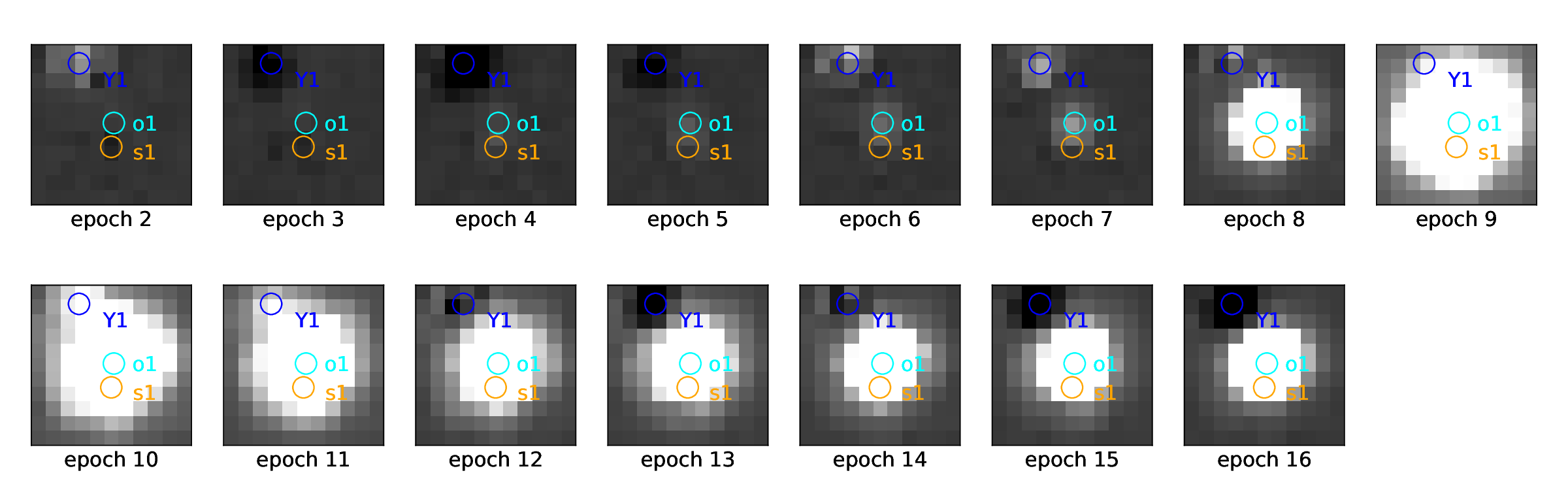}}
\end{minipage}
	\caption{30$^{\prime\prime}\times 30^{\prime\prime}$ difference images around \starget with the corresponding epoch 1 image as the reference \rev{in the W1 band}. \starget and the bright star are marked with o1 and s1, respectively. The young star (Y1), which is also variable, is located in the upper left corner. 
\label{fig:images}}
\end{figure*}

Fortunately, the region was covered by the Spitzer GLIMPSE 360 survey \citep{https://doi.org/10.26131/irsa205,2009PASP..121..213C,2003PASP..115..953B}. We retrieved calibrated images in the I1 and I2 bands, and indeed there is an infrared source (\target, hereafter \starget) 4\farcs5 north of S1 in these images (Figure \ref{fig:images}). These images were taken from 2013-11-06 to 2013-11-12, before the outburst. \starget has a brightness comparable to S1 in the $I2$ band, but is much fainter in the $I1$ band. The color of \starget is very red. We also examined the optical images at the Zwicky Transient Facility (ZTF), which were taken between March 2018 and November 2021, and no source was detected at the position of \starget.  

With a separation of 4\farcs5, the S1 source and \starget can be easily resolved in images. We retrieved a near-infrared image and photometry from UKDISS \citep{2007MNRAS.379.1599L} and found a weak source in the K band at the position of \starget (K=17.35$\pm$0.11 mag). The source appears extended in the K-band and is not detected in the J- and H-bands with upper limits of 20.0 mag and 19.2 mag, respectively. There is no significant variation of the source between the two observations made in 2007 and 2012. The region was observed by Herschel PACS \citep{https://doi.org/10.26131/irsa82} with blue and red filters on 2012-04-29.
A weak source is visible at the location of \starget in both images and is blended with another relatively bright source (Y1). We estimate the image flux by fitting two PSFs to a 9x9-pixel image centered on the two sources. The PSFs in the two bands were constructed by \citet{2011A&A...533A.119E}, and the centers of the PSF are fixed to the positions of the IRAC sources. \starget is detected marginally in the blue band at a signal-to-noise ratio (SNR) of $2.71$. 

\begin{figure*}
\figurenum{2}
\centering
\begin{minipage}{0.95\textwidth}
	\centerline{\includegraphics[width=1.0\textwidth]{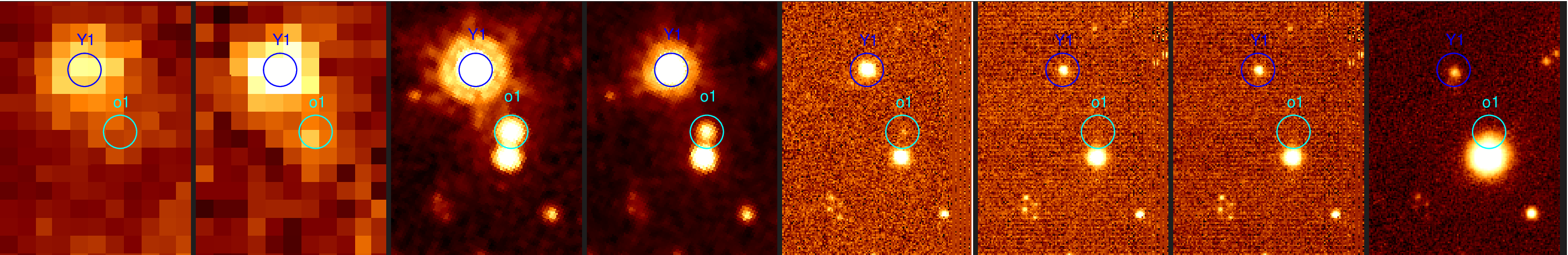}}
\end{minipage}
	\caption{Pre-eruption images (35$^{\prime\prime}$~$\times$~46$^{\prime\prime}$) of \starget in the infrared and optical. From left to right: Herschel 160 $\mu$m and 70$\mu$m, Spizter 4.5$\mu$m and 3.6$\mu$m, UKIDSS K, H \rev{and J}, PANSTARRs $i$. The green circle (3$^{\prime\prime}$ in radius) marks the position of \starget (o1), while the blue circle labels the nearby YSO (Y1). \starget is not visible in the image at wavelengths shorter than K-band. 
\label{fig:diff}}
\end{figure*}

\begin{figure*}
\figurenum{3}
\centering
\begin{minipage}{0.85\textwidth}
\centerline{\includegraphics[width=1.0\textwidth]{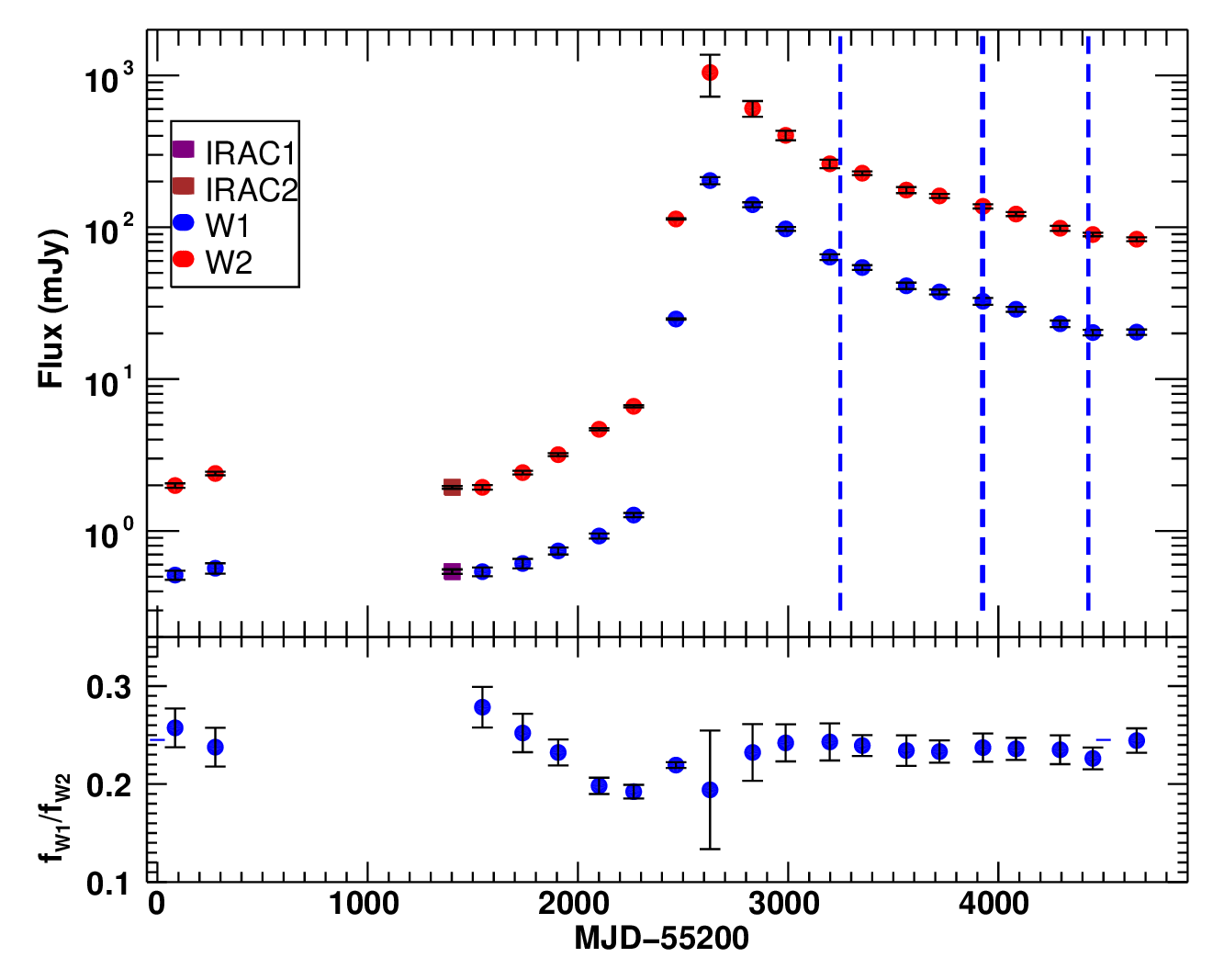}}
\end{minipage}
	\caption{In the upper panel, WISE survey mid-infrared light curves for \target in the W1 and W2 bands. Photometry is obtained from the differential image in the unWISE image with reference to epoch 1, except for the last four epochs (see the text for details). The magnitudes in the IRAC1 and IRAC2 bands at MJD = \rev{56602} are used for constant fluxes. Vertical dashed lines indicate dates of near-infrared spectroscopic observations. 
The bottom panel shows the flux ratios between the W1 and W2 bands over the period.
\label{fig:lightcurve}}
\end{figure*}

\starget is blended with S1 in the W1 and W2 images, and pre-eruption photometry from the WISE catalog dominates by S1. To obtain the complete MIR light curves of \starget, we downloaded the time-resolved unWISE coadds \citep{https://doi.org/10.26131/irsa524, 2018AJ....156...69M} and constructed differential images by subtracting the first-epoch WISE image from the rest of the images. The source has been visible in the different images since epoch 4, and the centroid coincides with \target (Figure \ref{fig:diff}), which supports that S1 did not vary significantly. We measured the differential magnitude of the source in the differential image using forced photometry. The light curves in the W1 and W2 bands were obtained by adding a constant, which was obtained from the Spitzer flux, to the difference fluxes \ref{fig:lightcurve}. Note that the seventh release of unWISE does not include images of the last four epochs, and instead we used the average magnitudes of NEOWISE single-exposure photometry. S1 contamination does not have a significant effect on our results because \starget was much brighter than S1 during the last four epochs.

An echelle spectrum was obtained using the Immersion GRating INfrared Spectrometer (IGRINS) mounted on the Discovery Channel Telescope at Lowell Observatory (now Lowell's Discovery Telescope) with an exposure of 4x1200 s on 27 November 2018, resulting in a spectrum with a resolution $\lambda/\Delta\lambda=45,000$ covering the infrared H and K bands \citep{2018SPIE10702E..0QM}. 
The magnitude measured from the acquisition image in the K band is approximately 12.5 mag, about 4.9 mag brighter than that in the UK hemisphere survey carried out in 20120314 or 20071029 \citep{2018MNRAS.473.5113D}. The IGRINS spectrum was reduced following a standard procedure described in \cite{2014SPIE.9147E..1DP} and \cite{2016zndo.....56067L}.  The SNR of the spectrum is very low in the H band but increases with wavelength, reaching $\sim$50 in the K band due to the steepness of SED.

\begin{figure*}
\figurenum{4}
\centering
\begin{minipage}{0.85\textwidth}
\centerline{\includegraphics[width=1.0\textwidth]{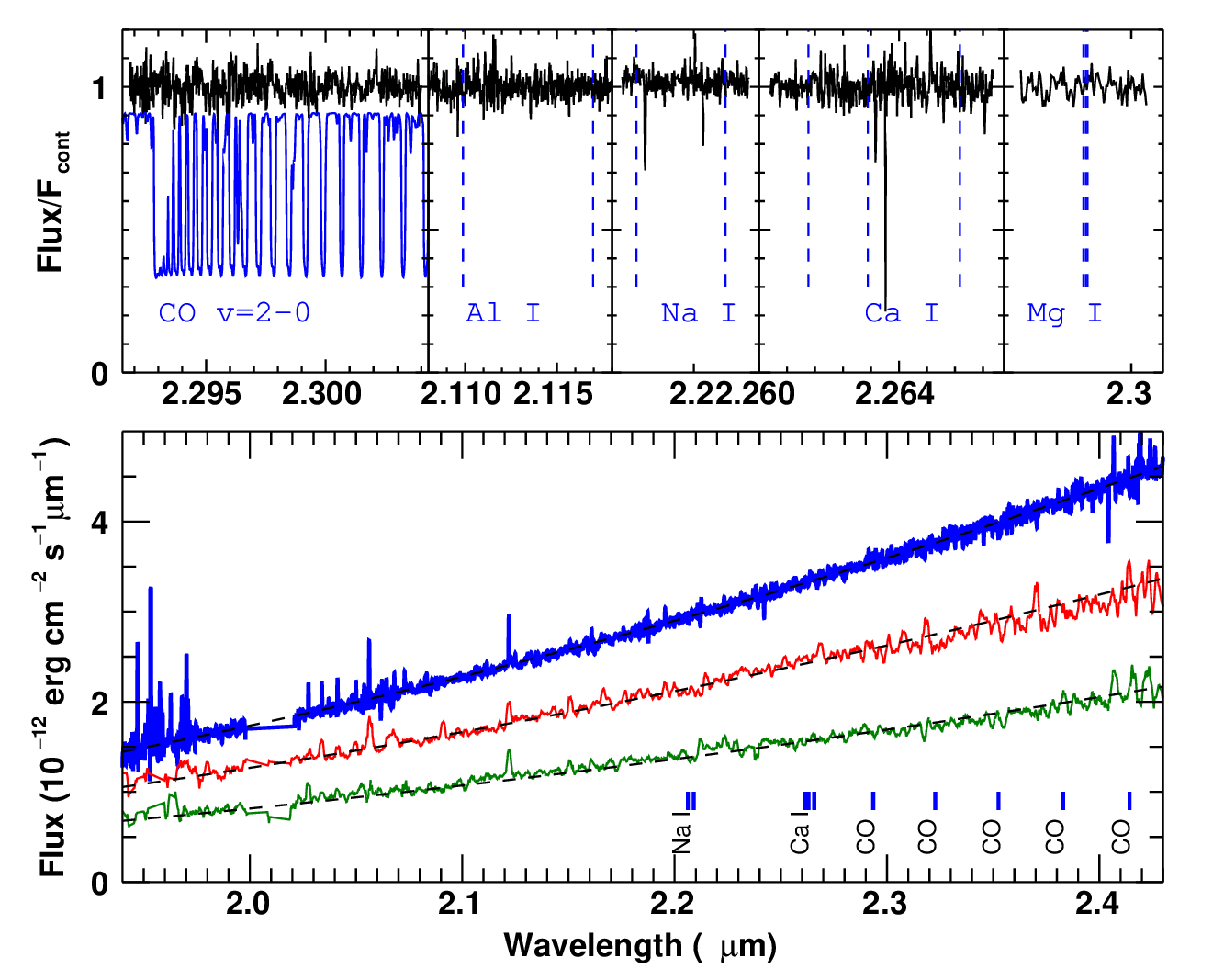}}
\end{minipage}
\caption{Near-infrared spectra of \target taken with IGRINS on Lowell Discovery Telescope (upper panel) and with the Triple Spectrograph on P200 (lower panel). For the sake of clarity, only spectra in the K band are shown. The IGRINS spectrum was smoothed by an 11-pixel boxcar and the TSpec by a 7-pixel wide boxcar. The upper panels show the normalized IGRINS spectrum around the CO $v=$2-1 overtune band, Al I, Na I, Ca I, and Mg I in the IGRINS spectrum. For reference, the CO first overtone band in the giant K star is shown in blue. The blue dashed vertical lines indicate the wavelengths of the lines at the systematic source velocity. The deep and narrow features around Na I and Ca I are due to the imperfect subtraction of telluric absorption lines and cosmic rays.
\label{fig:NIRspec}}
\end{figure*}

Three low-resolution IR spectra were obtained with the Triple Spectrograph (\citealp{2008SPIE.7014E..0XH}; hereafter TSpec) onboard the Palomar 200-inch telescope (P200) on 30 September 2020, 6 October 2020, and 18 February 2022, respectively. On-source exposure times were 120, 436, and 400 s in BAAB mode. TSpec simultaneously covers the J, H, and K band from 1-2.4~$\mu$m with a spectral resolution around 2700. Due to the steepness of SED, the SNR in the J band is less than 1, increases to about 3 in the H band and reaches up to 30 in the K band. The flux calibrated spectra are shown in Figure \ref{fig:NIRspec}. For the sake of clarity, the TSpec from September 30, 2020 is not shown because it is almost identical to the spectrum from October 6, 2020, but with slightly lower SNRs.

\section{Results} \label{sec:results}

\subsection{Environment of the source \label{sec:Eots}}

\begin{figure*}
\figurenum{5}
\centering
\begin{minipage}{0.95\textwidth}
\centerline{\includegraphics[width=1.0\textwidth]{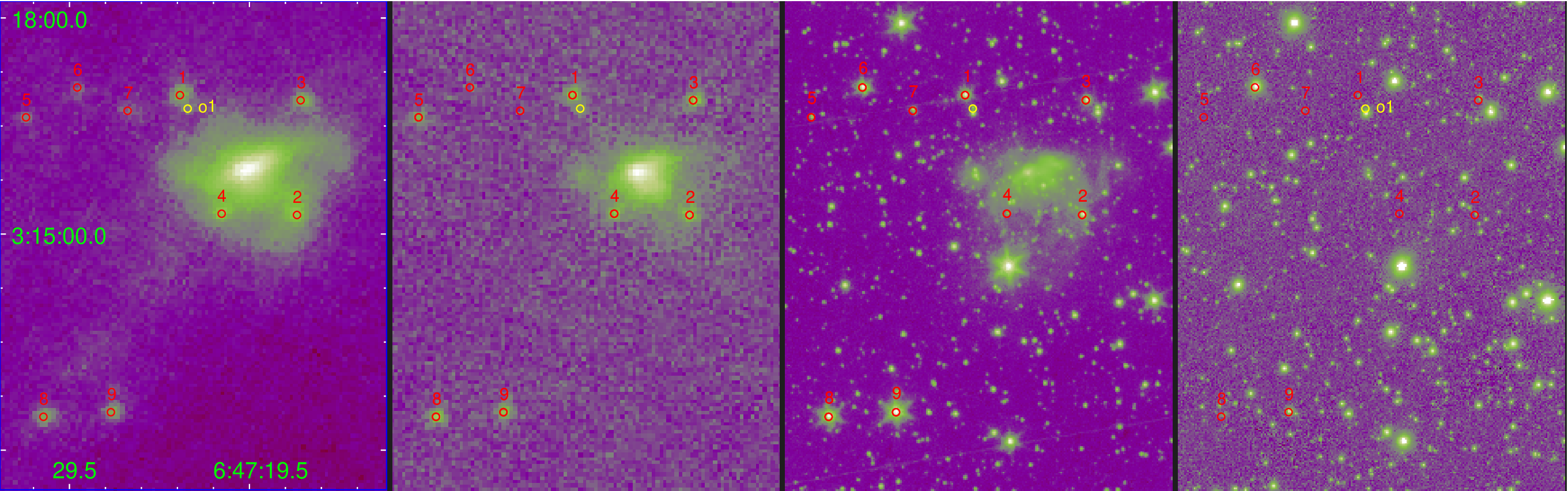}}
\end{minipage}
\caption{Nearby star formation in 5$\farcm$6 $\times$ 7$\farcm$2 images in the PACs red (160$\mu$m) and blue (70 $mu$m) bands of Herschel, IRAC 1 (4.5 $mu$m) of Spitzer, and $i$-band of pan-STARRS 
(panels from left to right). The positions of 9
YSO candidates are marked in red, and \starget is labeled o1 (yellow circle). 
\label{fig:environment}}
\end{figure*}

\begin{figure*}
\figurenum{6}
\centering
\begin{minipage}{0.85\textwidth}
\centerline{\includegraphics[width=1.0\textwidth]{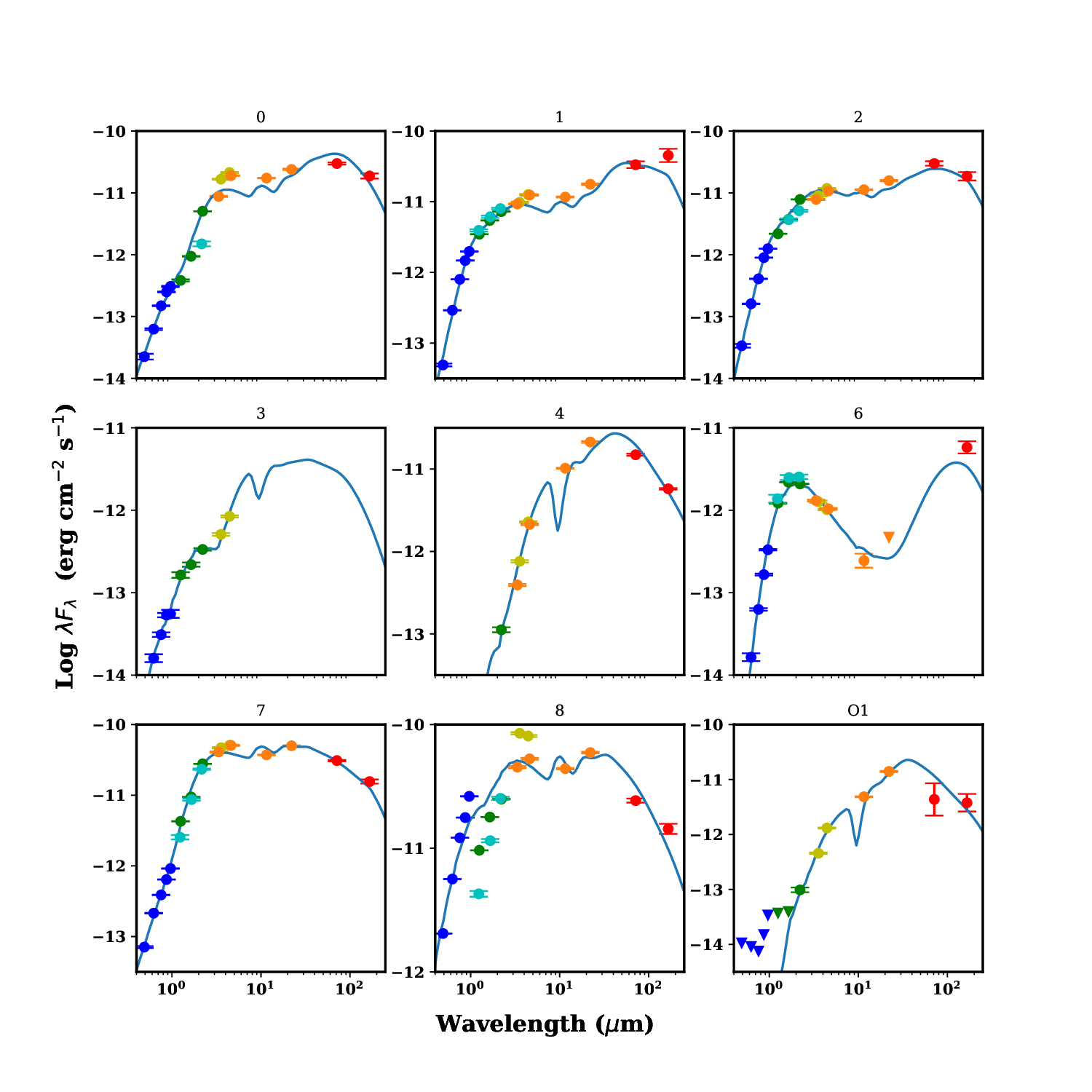}}
\end{minipage}
\caption{The broad band spectral energy distributions for nine YSOs marked in Figure \ref{fig:environment}. 
The sixth source (ID 5) is not shown because of the blending of two sources in the optical band. o1 is our source \starget. Non-simultaneous photometric data are collected from Pan-STARRS (blue), 2MASS (cyan), UKIDSS (green), IRAC of Spitzer (yellow), AllWISE data release (orange) and Herschel PACS (red). The large difference between 2MASS and UKIDSS data as well as the drop from PanSTARRS Y to J band in object 9 is due to variability rather than contamination. The best fitted YSO model is shown in a solid blue line.
\label{fig:YSOs}}
\end{figure*}

The main star formation region (SFR) near \starget is IRAS 0644+0319, which was identified as an intermediate mass SFR with a blob-shell morphology \citep{2014ApJ...784..111L}. Figure \ref{fig:environment} shows the 5$\farcm$6 $\times$ 7$\farcm$2 images in the Herschel's PACS red (160~$\mu$m) and blue (70~$\mu$m) bands, Spitzer's IRAC 1 (4.5~$\mu$m) and panSTARRS's $i$ band. IRAS 0644+0319 is seen in Figure \ref{fig:environment} as a large bright clump. \starget is located close to the outskirts of the SFR. Several point sources are visible on both the main SFR clump and the surroundings. They are likely young stellar objects, probably fragmented from the same parent molecular cloud. Stars 0, 2, 4, 5, 6 line up in a narrow strip in the middle panel of Figure 4, and the 160~$\mu$m image shows a weak tidal stream stretching from the main SF area to stars 7 and 8. We collect broadband photometric data for these nine sources marked in the second panel of Figure \ref{fig:environment} from the Herschel Archival Catalog, Spitzer Survey, Galactic Plane Survey of UKIDSS, and PanSTARRS archives \citep{2010A&A...518L...1P,2004ApJS..154....1W,2008MNRAS.391..136L,2016arXiv161205560C}. Their SEDs are plotted in Figure \ref{fig:YSOs}. The broadband spectral slope in the infrared suggests that they are candidate YSOs of Classes I and II. We also studied the MIR light curves of these objects from the WISE archive, and none showed periodic variations, which would have been the signature of dust-rich thermal pulsating AGB stars (TPAGB) \citep{2013A&A...560A..75K}.

\citet{2015ApJ...806...40L} calculated the kinematic distance of the SFR using the $^{13}$CO profile obtained with the Onsala 20 m telescope (OSO) in the direction of IRAS 06446+0319 \citep{2015ApJ...806...40L} with a beam size 35$\farcs$ FWHM. These authors found two peaks in the profile of the $^{13}$CO emission line separated by 2.05 km~s$^{-1}$, which was interpreted as two components of the molecular gas of column densities 3.66 and $2.99\times 10^{21}$~cm$^{-2}$ at distances 3.4 and 3.7 kpc, respectively, assuming that the molecular gas strictly follows the Galactic rotation curve with $V_\sun=240$~km~s$^{-1}$ and $R_\sun=8.34$~kpc. The small separation in two peaks of CO may be caused by bipolar outflows. In that case, the distance is likely between the two. Finally, we note that the VLSR velocity determined from $H_2$ lines in the infrared spectrum is about 40 km~s$^{-1}$, which is consistent with that from the CO measurement of the SFR, considering the uncertainty of wavelength calibration and possible warm $H_2$ outflows. This supports the idea that \starget is associated with SFR IRAS 06446+0319. We adopt a nominal distance of 3.5 with uncertainty of 0.2 kpc.  

\subsection{Spectral Energy Distribution in the Quiescent State and Class I YSO\label{sec:quiescent}}

\begin{figure*}
\figurenum{7}
\centering
\begin{minipage}{0.85\textwidth}
\centerline{\includegraphics[width=1.0\textwidth]{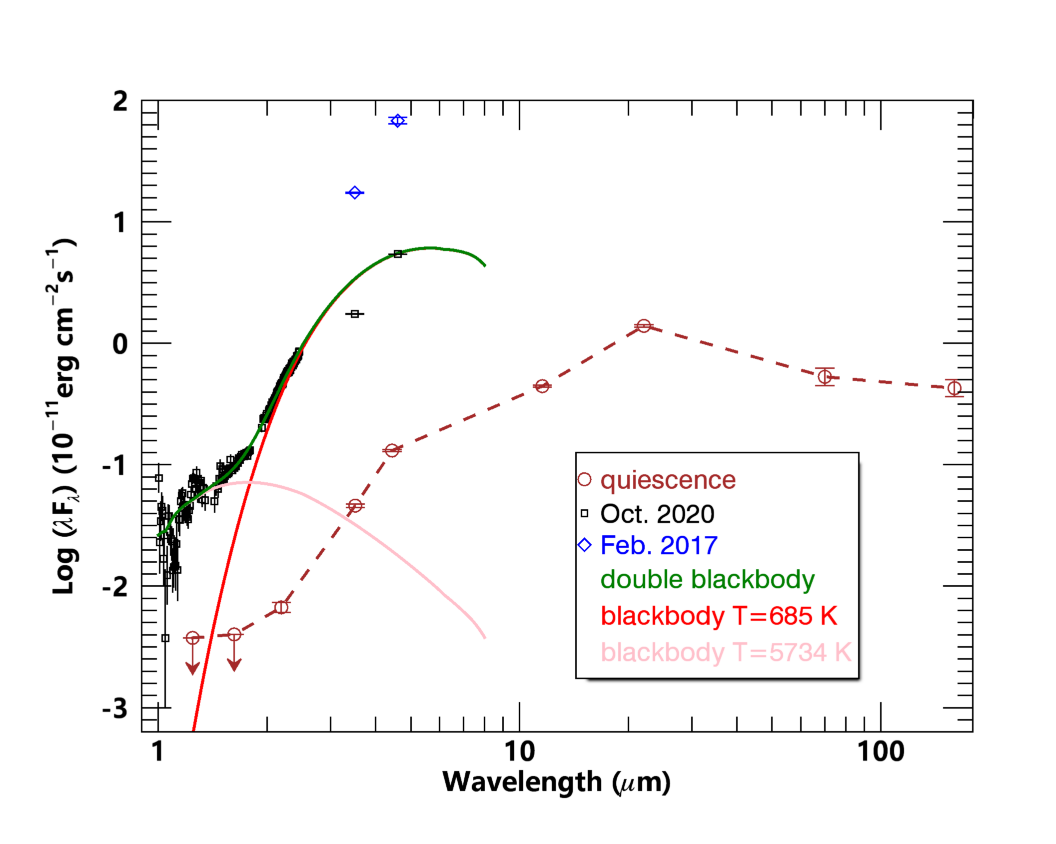}}
\end{minipage}
\caption{The spectral energy distribution of \starget in the quiescence (orange circles connected by dashed lines). For comparison, the binned TSpec of October 6, 2020 and quasi-simultaneous NEOWISE fluxes are shown in black squares, and the NEOWISE fluxes at the peak of the eruption in 2017 are also plotted (blue diamonds). Note that Tspec below 1.5 $\mu$m may be affected by systematic background subtraction. The best-fit model to TSpec and its constituted components are shown as solid lines.   
\label{fig:J0647sed}}
\end{figure*}

Photometric data before the outburst are plotted in Figure \ref{fig:J0647sed}. Although these data are collected at different times, the SED represents the typical quiescent emission of \starget. K-magnitudes obtained in 2007 and 2012 were not significantly different. Little can be seen in the WISE differential images for epoch 1-epoch 0 and epoch 2-epoch 0, suggesting that the source stayed in a quiescent state before epoch 3. The Spitzer and Herschel observations were all performed before epoch 3. S1 and \starget are mixed in the WISE W3 and W4 bands. Since S1 does not show an infrared excess in W1 and W2, an infrared excess in the W3 or W4 bands is unlikely. S1 is not variable in the W1 or W2 bands, and no systematic residuals are seen in the star position in differential images. Thus, we estimated the magnitude of W3 and W4 of S1 by fitting a
stellar model to its SED in the optical and near-infrared (NIR) as well as Spitzer I1 and I2. The SED is well fitted with an ATLAS9 model. Extrapolating S1 photospheric emission suggests that it contributes $\sim$1.3\% of the observed W3 flux and $\sim$0.1\% of the W4 flux. 

The SED of \starget in the quiescent state increases steeply from 3 to 20 $\mu$m and then flattens to the far infrared. Quiescent WISE colors reside at the location of Class I YSOs on the color-color diagram, far from the AGB star locus (Figure 5 and the second panel of \citealp{2014ApJ...791..131K}; also \citealp{2017MNRAS.472.2990L}). With a spectral slope of $n_{4.5-24}=d\log (\lambda  f_\lambda)/d\log\lambda\simeq 1.47$ and an upper limit of bolometric temperature \citep{1993ApJ...413L..47M} of 165~K, it is classified as Class I YSO \citep[Figure 3 of][]{2023ApJS..266...32P}. However, the upper limit of the bolometric temperature is near the boundary between Class I and Class 0, so Class 0 cannot be completely rejected. 
The SED is reasonably well fitted by the YSO models in Figure \ref{fig:YSOs} \citep{2017A&A...600A..11R}.  
Bolometric flux is estimated to be $2.5\times10^{-11}$ ~erg~cm$^{-2}$~s$^{-1}$ by integrating the SED from the near to the far infrared with a simple log-linear interpolation. This number is within a factor of two of the integration of the best-fitted YSO model. 
The bolometric correction for the luminosity of W2 is $C=L_{bol}/\nu L_\nu(W2)\simeq 20$. With the distance in \S \ref{sec:Eots}, we derived a quiescent luminosity of $9(d/3.5$kpc$)^2~L_\sun$. This luminosity and the absence of periodic variations eliminate the possibility of \starget being an embedded AGB interloper.

\subsection{Mid-infrared Outburst and Evolution of SED\label{sec:outburst}}

The W1 and W2 light curves measured from differential images, combined with binned NEOWISE photometry for the last 4 epochs, are shown in Figure \ref{fig:lightcurve} with the constant fluxes estimated from the IRAC observations, which are close to epoch-3 of the WISE observations. \starget brightened gradually from 2014-03-29 to 2016-03-21 before the giant eruption of a factor of 100 in a year, then faded by a factor of five until the end of 2021. This gives a fading rate about 0.5~mag~yr$^{-1}$. With a total amplitude of about 500 times in the $W2$ band, it is the second highest amplitude infrared outburst discovered in a YSO after WISE~J142238.82-611553.7 (WISEA1422 for short \citealp{2020MNRAS.499.1805L}).

To explore the evolution of the spectral energy distribution (SED), we show the SED of the quiescent state, TSpec taken in October 2020 and quasi-simultaneous WISE magnitudes, and W1 and W2 in the peak of the light curve in Figure \ref{fig:lightcurve}. Because there are no apparent photospheric absorption features in the NIR spectrum, we fit the NIR spectrum with a double black-body model. The best fit converges to a cold component with $T_{BB}=685\pm 78$K and $f_{BB}=(1.03\pm 0.04)\times 10^{-10}$~erg~cm$^{-2}$~s$^{-1}$ ($L_{BB}=(39\pm 1)(d/3.5{\mathrm kpc})^{-2}~L_\sun$) with $\chi^2$ ($\chi^2/dof=4545/6452$). A review of normalized residuals suggested that the low $\chi^2$ is due to an overestimation of the error in the very low SNR spectrum portion. Consequently, the uncertainties of the parameters are overestimated. However, the parameters of the hotter component and the extinction were poorly constrained because of their strong coupling. The extrapolation of the best-fit model to MIR results in a flux of $W2$ consistent with the observed flux, but overpredicts the flux of $W1$ (Figure \ref{fig:J0647sed}). We speculate that the deficit in W1 flux may be explained by iced molecular absorption, since strong iced absorption has been observed in embedded YSOs of Class 0 to 2 by Spitzer and Akari \citep[e.g.]{2008ApJ...678..985B,2013ApJ...775...85N}. Further spectroscopic observation in the mid-infrared can test this.

The lack of photospheric absorption properties may be due to low SNR spectra in the J- and H-bands and the dominance of the low-temperature blackbody in the K-band. To test the robustness of the parameters of the latter component, we also fitted the spectrum with a blackbody plus a stellar template from the Extended IRTF stellar library \citep{2017ApJS..230...23V} due to its spectral resolution similar to that of TSpec ($R\simeq 2000$). Various stellar templates of the types G8 to M5 lead to fittings with a similar $\chi^2$, the parameters of blackbody ($T_{BB}=590-613$~K, $L_{BB}=(43-47)~L_{sun}$) to the double blackbody models.


The MIR color is quite stable during the outburst with $R=f_{W1}/f_{W2}\simeq 0.2-0.24$ (Figure \ref{fig:lightcurve}) despite dramatic flux variations. Three near-infrared spectra taken in December 2018, October 2020, and February 2022 are consistent with the same shape and only the brightness varied (Figure \ref{fig:NIRspec}). Furthermore, the brightening of 4.9 mag in the K-band acquisition image in December 2018 with respect to the quiescent state was similar to the difference in $ W1$ or $ W2$ of the same period compared to the quiescent state. This result indicates a constant slope of the IR SED, in contrast to other FUors and EXors during outbursts, whose SEDs in the IR usually change shape as sources vary (see \S  \ref{sec:discussion}). Direct scaling of the black-body model to match the peak $W2$ flux gives an estimate of the black-body luminosity 400$(d/3.5{\mathrm kpc})^{-2} ~L_\sun$ at the peak of outburst.  
However, if the shape of the SED remains in all bands the same as that in the quiescent state, the bolometric luminosity would be 4650$(d/3.5{\mathrm kpc})^{-2}~ L_\sun$. In many YSOs, the variability amplitude decreases from the mid-infrared to the submillimeter band \citep{2019MNRAS.487.4465M,2020MNRAS.495.3614C,2013ApJ...776...77K}, so the real peak luminosity is likely between the two. With the luminosity and temperature of the black body, we can estimate a minimum size of $>$6.5$(d/3.5{\mathrm kpc})$~AU for the emission region during the peak outburst using the black-body luminosity 400$L_{\sun}$. 

\subsection{Estimate of the mass accretion rate}

The main source of energy for the outburst is the gas accretion onto the YSO, although the detailed physical process that caused such a sudden mass inflow is not yet clear. The accretion power can be written as
\begin{equation}
	L_{acc}\simeq\beta\frac{GM_*\dot{M}}{R_*}=
	312\beta\left(\frac{M_*}{M_\sun}\right)\left(\frac{R_*}{R_\sun}\right)^{-1}\dot{M}_{-5}L_\sun
\end{equation}
where $\beta<1$ is the radiation efficiency, $\dot{M}_{-5}$ is the mass accretion rate in units of $10^{-5} M_\sun$~yr$^{-1}$, and $M_*$ and $R_*$ are the mass and radius of the YSO, respectively. We will adopt the mean $R_*/M_* = 3.9 R_\sun/M_\sun$ with a scatter of 1.9 for TTS \citep{2020MNRAS.499.1805L}. At a high accretion rate, we might underestimate $R_* / M_*$ due to a possible expansion of the upper layer of the star caused by the advection of the disk energy \citep{2012ApJ...756..118B}. Taking into account the uncertainty in the bolometric correction, $M_*/R_*$ and $\beta$, a mass accretion rate of at least a few $10^{-5} M_\odot yr^{-1}$ is required to explain the observed luminosity. This is well within the range for FUors. 

\subsection{Emission and Absorption lines in K-band \label{sec:eml}}

\begin{figure*}
\figurenum{8}
\centering
\begin{minipage}{0.85\textwidth}
\centerline{\includegraphics[width=1.0\textwidth]{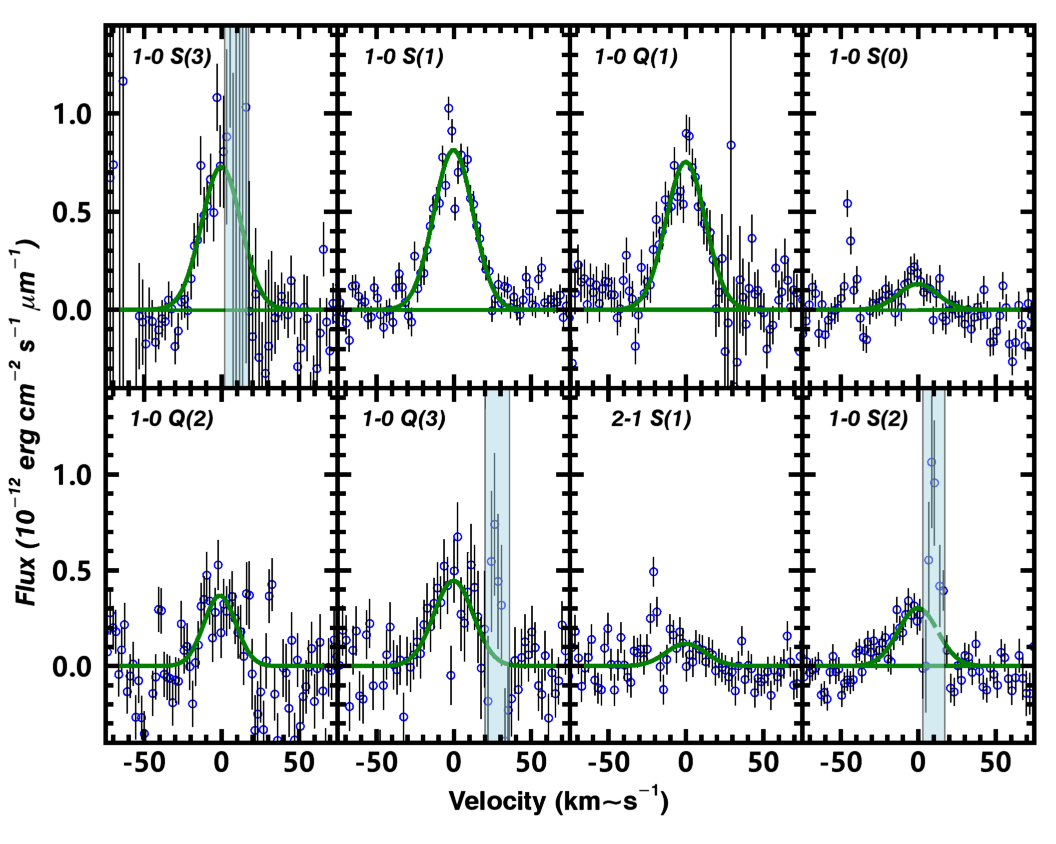}}
\end{minipage}
\caption{The profiles of $H_2$ lines from the IGRINS spectrum. Regions strongly affected by OH absorption lines are masked by cyan bars.  
\label{fig:H2}}
\end{figure*}
The SNRs obtained in the J and H bands are low, and so we will focus on the K-band spectra. Several $H_2$ emission lines, 1-0 S (1), 1-0 S (2), 1-0 S (3), and 1-0 Q (1), are clearly visible in the IGRINS spectrum at 1.95756, 2.03376, 2.12183 and 2.40659~$\mu$ m. Additional $H_2$ lines 1-0 Q(2) (2.41344~$\mu$m), 1-0 Q(3) (2.42373~$\mu$m), 2-1 S(1) (2.24772~$\mu$m) and 1-0 S(0) (2.22329~$\mu$m) are also present. The 1-0 S(1), 1-0 S(2) and 1-0 S(3) lines are also seen in the P200 spectra. These lines are single-peaked and can be fitted by a Gaussian function with $v=18.9\pm0.3$ and $\sigma=12.9\pm 0.3$~km~s$^{-1}$ in the IGRINS spectrum (see Figure \ref{fig:H2}). To detect weak lines, we have tied the centroid and width for all lines. $H_2$ is a factor of 2-4 narrower than those in classical T-Tauri stars \citep{2007ApJ...670L..33T} but broader than emission from the outer disks in Herbig Ae/Be stars \citep{2011A&A...533A..39C}. The line width suggests that it comes from the circumstellar material, rather than from the interstellar medium. Taking into account the motion of the Sun and Earth, the velocity with respect to the LSR is 40.1~ km~s$^{-1}$, which is consistent with
the LSR velocity of CO lines from IRAS 06446+0319 \citep{2015ApJ...806...40L} considering all uncertainties. 
The measured line fluxes are listed in Table \ref{tab:eml}. The weak lines 1-0 S (0) and 2-1 S (1) are sensitive to the exact subtraction of the continuum, which is affected by telluric OH absorption lines. Their flux errors can be as large as 50\%. No other emission or absorption lines are seen. Neither recombination lines Br$\gamma$ at 2.1661~$\mu$m and He I nor CO lines are seen in absorption or emission. Photospheric metal absorption lines such as from Na I and Ca I, commonly seen in the IR spectra of low-mass YSO (e.g., \citealp{2012PASP..124.1137F}), are not seen in the IGRINS spectrum. 

$H_2$ 1-0 Q(3) and 1-0 S(1) share the same upper level, as do 1-0 Q (2) and 1-0 S (0). The ratios of these lines can be used to estimate their extinction. Due to the weakness of 1-0 S(0), we will not use the latter line ratio. The observed ratio $1-0\ Q(3)/1-0\ S(1)=0.77\pm 0.09$ is consistent with the theoretical value 0.70 \citep{1977ApJS...35..281T}. Line ratios of H2 1–0 S (1) to 2–1 S (1) and 3–2 S (3) have traditionally been used to diagnose the excitation mechanism of H2 molecules. In \starget, 1–0 S (1)/2–1 S (1)$=6.3\pm 1.3$ is higher than the value 2-3 expected for UV excitation, but is consistent with shock excitation \citep{1995ApJ...455..133H}. The presence of narrow $H2$ emission lines is also consistent with a YSO. 

$H_2$ emission lines are considered as tracers of outflows in young stellar objects, and were detected in only 3 of 9 FUors with $SNR>3\sigma$ (also V346 Nor, \citealp{2020ApJ...889..148K}; V960 Mon, \citealp{2020ApJ...900...36P} ) and a few EXors (\citealp{2011ApJ...736...72K};\citealp{2019AJ....158..241H,2020AJ....160..164H}; V899 Mon, \citealp{2021ApJ...923..171P}). Most of these detections are made only in 2.218$\mu$m transition with the exception of two objects. When high-resolution spectra are available, the line widths are in the range 20-40 km~s$^{-1}$, much narrower than hydrogen lines or other metal emission lines whenever observed, consistent with being a large-scale outflow origin, rather than high velocity outflows from the inner disks (e.g. \citealp{2009AJ....138..448H}). 


\section{Discussion\label{sec:discussion}}


We have argued that \starget is a deeply embedded Class I protostar according to the pre-outburst SED, location at the edge of an SFR, consistency of $H_2$ velocity with $CO_2$ of the SFR, low quiescent luminosity, and absence of periodic fluctuations. The distance of the SFR, around 3.5-3.7 kpc, derived from the CO kinematic measurement, is also consistent with the constraint of extinction of foreground stars with known distances. The quiescent luminosity of $9~L_\sun$ obtained in \S\ref{sec:quiescent}, which is close to that of V1647 Ori \citep{2008AJ....135..423A}, places it in the upper quarter of the Class I YSOs in the Spitzer \& Herschel survey samples \citep{2016ApJS..224....5F, 2023ApJS..266...32P}.  
If the accretion power dominates the total radiative output in the Class I phase, the bolometric luminosity provides an estimate of the quiescent mass accretion rate ($\dot{M}\simeq LR/GM$) to be a few $10^{-6}$~$M_\sun$ year$^{-1}$ for the typical mass and radius of a protostar. However, if the radiation is mainly powered by the protostar, the mass of the star is estimated to around 0.7-1.3~M$_\sun$ for a stellar of an age less than $10^7$ years according to the models of \citet{2000A&A...358..593S}. A lower limit on stellar mass of 0.58 M$_\sun$ is derived from the luminosity evolution models for exponentially declining infall rates \citep{2017ApJ...840...69F}. 

\starget exhibited a 500-fold increase in MIR brightness over a two-year period, followed by a slow decline. This eruption amplitude is the second largest recorded among all known YSO eruptions in the MIR \citep{2020MNRAS.499.1805L}. The tenfold decay within six years is much slower than that of EXor eruptions but considerably faster than FUor eruptions. From the light curve alone, \starget can be classified as an intermediate-type eruption with an exceptional amplitude such as WISEA1422. 

Infrared-only eruptions have been observed in several young stellar objects (YSOs) over the past decade, such as OO Ser \citep{2007A&A...470..211K} and WISEA1422. The most extreme example of this type is Hops 383, a Class 0 object, which had a 35-fold eruptive amplitude at 24 $\mu$m \citep{2015ApJ...800L...5S} and was not detected in the J, H, and K bands up to 23.1, 21.8, and 20.4 magnitudes, respectively \citep{2017ATel.9969....1F}. Other sources without detection in the J and H bands were reported in \citet{2017A&A...603A..26N, 2023MNRAS.522.2171N}. These sources, together with \starget, all had eruptions of intermediate duration, which is consistent with the finding that intermediate-type eruptions are the most commonly seen type in near-infrared time-domain surveys \citep{2017A&A...603A..26N}.  

\starget does not display detectable stellar photospheric absorption lines or CO absorption bandheads of a typical M supergiant in the near-infrared spectrum during the outburst, which is the benchmark for classical FU Orion objects \citep{2008AJ....135.1421G, 2018ApJ...861..145C}. Neither shows Br$\gamma$ or Pachen emission lines or CO bandheads in emission that are usually seen in EXor eruptions. Featureless spectra have been reported in a handful of YSO outbursts so far. OO Ser is the first object of this type \citep{2007A&A...470..211K}. The group includes three other objects \footnote{Gaia 19bey has switched to a featureless spectrum toward the end of the outburst while showing a spectrum for an EXor in the early outburst \citep{2020AJ....160..164H}. The outburst of HBC 494\citep{2015ApJ...805...54C} was indirectly inferred from the brightening of the nebula. They appear different.}: 2MASS 22352345+7515076\citep{2019MNRAS.483.4424K}, WISEA 1422\citep{2020MNRAS.499.1805L}, and VVVv815\citep{2017MNRAS.465.3039C}. All of them are deeply embedded sources.

Another interesting feature is that the $W1-W2$ color stays almost constant from the pre-outburst to the outburst at \starget. The constant shape can extend to the NIR band by comparing the NIR spectra obtained at different epochs. We can compare this with other YSOs in outbursts. Two FUors Gaia 17bpi and Gaia 18dvy showed a pattern of bluer when brighter in the MIR band during an eruption with an amplitude $>6$ mag in the $W1$ bands \citep{2018ApJ...869..146H}. WISEA1422 becomes redder from quiescence to eruption and, with a short duration of turning blue after the peak, becomes even redder afterwards \citep{2020MNRAS.499.1805L}. The V1647 Ori analogy ASASSN-13db has a color variation very similar to WISEA1422. However, a wavelength-independent decay was observed for the OO Ser outburst between 1996 and 2005, 
while the pre-outburst mid-infrared color is not available \citep{2007A&A...470..211K}.

In summary, \starget is a deeply embedded young stellar object that has four distinct features: a medium-length eruption, an unusually high outburst intensity in the mid-infrared, a featureless K-band spectrum, and wavelength-independent changes in the near-infrared and mid-infrared bands. Apart from the last one, it is similar to WISEA1422. We will show that these peculiarities can be explained by a dusty disk outflow or an inflated outer disk, which also transforms primary radiation from stars and disks into longer infrared wavelengths.

CO absorption bandheads in an outburst FUor are formed in the hot inner accretion disk \citep{2018ApJ...861..145C,2022ApJ...936..152L}, whose inner radius is pushed by high gas pressure to the surface of YSO when the accretion rate is high. The Br$\gamma$ and Paschen emission lines in EXors are produced by gas photo-ionized by UV photons, generated in the shock of the magnetospheric accretion column for a low to moderate accretion rate \citep{2021MNRAS.504..830G}. In V899 mon, whose properties lie between FUor and EXor, CO absorption lines were observed in the bright outburst phase \citep{2015salt.confE..69N}, while CO emission lines were detected when the source faded \citep{2021ApJ...920..132P}, supporting this idea. Photospheric absorption lines are absent in \starget because the YSO and the inner accretion disk are completely obscured. Recombination lines are not observed because the emission line region is obscured as well or because the accretion rate is high, so magnetospheric accretion is suppressed as in the case of a FUor. The high accretion rate in this object is supported by its high outburst luminosity.

The common trait of spectroscopically featureless outbursts is that the sources are deeply embedded. This is consistent with the above scenario. 
The obscurer could be the inflated outer disk, a dusty wind, or even dust in the envelope when the system is viewed edge-on. This same obscurer may also absorb the radiation from the inner accretion disk and the central star and reprocess it into mid-infrared emission with a low effective temperature. As estimated in Section \ref{sec:outburst}, the MIR emission region has a size of at least 6.5AU, which is in the range of protoplanetary disks around Class I stars \citep{2020ApJ...890..130T}, but much smaller than the envelope. The high luminosity in the infrared suggests that the reprocessor should have a large covering factor. The constant observed W1-W2 in \starget implies that the reprocessing region expands (shrinks) as the source brightens (dims) because it is expected that the temperature of the dust will vary with the luminosity of $T\propto L^{1/4}$ in a steady structure. This indicates that the obscurer is more likely a disc wind, as the time scale to adjust for the outer disk should be long. We hypothesize that this is the result of the increasing launching radius of the dusty outflow with the increase of disk luminosity.

\citet{2020MNRAS.499.1805L} proposed a second possibility that the eruption of WISEA J142238.82$-$611553.7 could be caused by the fragmentation of the disk or the falling material from the envelope.
Because the specific gravitational binding energy in situ (a size of $>$ 6.5AU) is so small, it requires an unrealistically high mass infall rate ($\dot{M}\sim LR/GM \sim 10^1$ $M_\sun$~ year$^{-1}$) to explain the high luminosity and large emission size for \starget. The total mass accreted during the outburst far exceeds the range for the disk or envelope in Class 1 YSOs. Therefore, the in situ scenario can be rejected here.

In summary, we anticipate the dust-obscuring and reprocessing scenario to explain the size of the emission region of the IR outburst, the featureless K-band spectrum, and the constant IR colors. To confirm the feasibility of this, further work is needed, including detailed radiation transfer modeling and numerical simulations of disk winds under extreme changes in the accretion rate of the inner disks.

\section{Conclusion} \label{sec:conclusion}

We discovered a long-lasting and massive outburst in \starget, which was only detected at wavelengths longer than the K band when quiescent. The source was identified as an embedded Class I YSO from its association with a SFR, the SED, and the bolometric luminosity in quiescence. Its near-infrared spectrum is different from that of classical FUors, EXors, or many other known intermediate-type outbursts in YSO due to its lack of absorption or emission lines other than $H_2$. Its SED during the outburst was dominated by MIR emission with an effective temperature of 600-700 K and a weak near-infrared excess component. We interpret the observed properties in the context of an obscured accretion outburst. Both the YSO and the inner accretion disk in \starget are obscured by a thick dusty outflow or cold outer disk/envelope, which reprocesses the stellar and disk emission into mid-infrared light. The weak near-infrared excess is probably the scattered accretion disk emission. The size of the reprocessing emission region is $>$6.5 AU. As a result, photospheric absorption lines from stellar and disk photospheres are not directly seen. We estimate a maximum mass accretion rate of the outburst of at least a few $10^{-5}$ $M_\sun$~yr $^{-1}$. The fading rate (0.5 mag yr$^{-1}$) is steeper than FUors, but much shallower than EXors.  
The variability amplitude in the MIR is only one magnitude smaller than the outburst in another serendipitously discovered WISEA1422 \citep{2020MNRAS.499.1805L}, whose NIR spectrum is also similar to \starget with a nearly featureless continuum with the exception of the $H_2$ emission line, but the spectrum was taken when the source was 6 magnitude below the peak. This source is also identified as an embedded Class I YSO. The color changes in the two sources are different. The color of W1-W2 remains nearly constant in \starget during the outburst, while it changes dramatically in WISEA1422 and outbursts in other YSOs. This may indicate the presence of a population of heavily embedded outbursts. The true number of such sources and their contribution to our understanding of the mass growth of YSOs have yet to be fully explored.

\begin{acknowledgements} 

The authors thank Dr. Gregory Herczeg for thorough reading of the first manuscript version and constructive suggestions. We thank the referee for helpful comments. This work is supported by NSFC funding (NSFC-11833007), the China Manned Space Project (NO.CMS-CSST-2021-A13), the Ministry of Science and Technology (2022SKA0130102), and the Cyrus Chun Ying Tang Foundations.

This work used the Immersion Grating Infrared Spectrometer (IGRINS) that was developed under a collaboration between the University of Texas at Austin and the Korea Astronomy and Space Science Institute (KASI) with the financial support of the Mt. Cuba Astronomical Foundation, the US National Science Foundation under grants AST-1229522 and AST-1702267, McDonald Observatory of the University of Texas at Austin, the Korean GMT Project of KASI, and Gemini Observatory. These results were obtained using the Lowell Discovery Telescope (LDT) at Lowell Observatory.  Lowell Observatory is a private, non-profit institution dedicated to astrophysical research and public appreciation of astronomy and operates the LDT in partnership with Boston University, the University of Maryland, the University of Toledo, Northern Arizona University, and Yale University. TSpecs are obtained through the Chinese TAP program.

This research has used the NASA/IPAC Infrared Science Archive, which is funded by the National Aeronautics and Space Administration and operated by the California Institute of Technology. "

The Pan-STARRS1 (PS1) Surveys have been made possible through contributions of the Institute for Astronomy, the University of Hawaii, the Pan-STARRS Project Office, the Max-Planck Society and its participating institutes, the Max-Planck Institute for Astronomy, Heidelberg, and the Max-Planck Institute for Extraterrestrial Physics, Garching, The Johns Hopkins University, Durham University, the University of Edinburgh, Queen's University Belfast, the Harvard–Smithsonian Center for Astrophysics, the Las Cumbres Observatory Global Telescope Network Incorporated, the National Central University of Taiwan, the Space Telescope Science Institute, the National Aeronautics and Space Administration under Grant No. NNX08AR22G issued through the Planetary Science Division of the NASA Science Mission Directorate, the National Science Foundation under Grant No. AST-1238877, and the University of Maryland.
\end{acknowledgements}

\facilities{IRSA, Spitzer, WISE, Herschel, P200, LDT, ZTF}

\begin{deluxetable*}{lll}
	\tablenum{1}
	\tablecolumns{3}
	\tablewidth{0pt}
	\tablecaption{Flux of Emission Lines from IGRINS\label{tab:eml}} 
	\tablehead{\colhead{Line ID} & \colhead{Wavelength} & \colhead{Intensity}\\
		\colhead{}& \colhead{$\mu$m} &\colhead{$10^{-16}$\ erg~s$^{-1}$}}
	\startdata
	1-0 S(0) & 2.22329  & 0.37$\pm$0.05\\
	1-0 S(1) & 2.12183  & 1.89$\pm$0.05\\
	1-0 S(2) & 2.03376  & 0.68$\pm$0.05\\
	1-0 S(3) & 1.95756  & 1.34$\pm$0.13\\ 
	1-0 Q(1) & 2.40659  & 1.98$\pm$0.08\\
	1-0 Q(2) & 2.41344  & 0.75$\pm$0.11\\
	1-0 Q(3) & 2.42373  & 1.12$\pm$0.13\\
	2-1 S(1) & 2.24772  & 0.30$\pm$0.06\\
	\enddata
\end{deluxetable*}

\bibliography{reference}{}
\bibliographystyle{aasjournal}

\end{document}